# Overview of the development of smart classrooms under information technology: development and innovation of hardware and software

**Yanying Cheng**

*College of Business and Management, Jilin University, Zhuhai, 511500, China,*
*948837110@qq.com*

*Abstract:* With the rapid development of information and communication technology (ICT), smart classroom has become an important trend in education modernization. This article reviews the development of smart classrooms from the hardware and software levels. The hardware describes the transformation from the construction of basic ICT facilities in single mode to a multi-modal information cloud platform. In terms of software, we look at the evolution of related supporting algorithms and technologies from the platform construction technology to the integration of advanced artificial intelligence (AI) technology from the perspectives of learning analysis and data mining. Provide guidance and suggestions for future educators, researchers and policymakers on the future direction of smart classrooms.

## 1.    Introduction (Heading 1)

The rapid development of Information and Communication Technology (ICT) has become a vital force in facilitating individual learning. ICT refers to various technological tools and resources used for communicating, creating, disseminating, storing, and managing information. It's acknowledged that by 2022, the goal is to have digital campus constructions covering all schools in China, indicating a direction towards further informatization and intelligentization of education [1]. Smart classrooms, as a sophisticated form of digital education construction, are recognized as a trend in China's educational future development. They represent a new and improved way of teaching that combines ICT with education, redefining the space for teachers and students to restructure learning activities.[2] Tools like electronic blackboards and educational systems provide data on learning progress and outcomes, enabling teachers to tailor courses based on student needs. Instant feedback systems allow for a clear understanding of learning progress [3]. Students also engage with information technology tools independently, fostering observation and creativity.[4]

The concept of artificial intelligence (AI) has risen swiftly in recent years. AI involves the development of machines with the ability to perform functions similar to human cognition, learning, decision-making, and environmental adaptation. AI is already making strides across various fields, such as autonomous driving technologies [5] and medicine [6]. In the realm of education, the evolution from simply using online materials for study to the inclusion of smart, adaptive learning for personalized education is notable [7]. AI applications in education have expanded to teaching, management, and tutoring, offering intelligent learning experiences through innovative virtual systems and predictive data analytic. Artificial intelligence-assisted education includes intelligent education, innovative virtual learning, data analysis and prediction, etc. The main scenarios and technologies of artificial intelligence in education are shown in Table 1. With the advancement of educational modernization and the rising requirements of the new era, artificial intelligence education is playing an increasingly important role. Intelligent education systems provide teachers and learners with timely, personalized guidance and feedback. Thereby improving the learning value and efficiency.

With the continuous upgrading of educational informatization and changes in the teaching environment, the teaching forms of smart classrooms are also iterating, and the direction of research targets has also changed. Figure 1 shows the results of the year-by-year assessment of technology patterns explored by Kaur et al [8]. Early (pre-2016) research on smart classroom-based instructional design focused primarily on the updating and designing of instructional methods through iteration of information technology equipment. Kristopher Scott and others believe that devices installed in smart classrooms can clearly identify students' location and progress and determine current learning activities [9]. Liu Bangji (2016) focused on smart classroom research and implementation strategies, emphasizing the process structure, teaching goals and teaching design of smart classrooms, and proposed corresponding teaching strategies [10]. Byan Jinjin (2016) analyzed the effectiveness of smart classroom learning models from the perspective of learning model design [11]. After 2016, smart teaching often involves the empowerment of artificial intelligence. These studies have shifted from focusing on hardware devices to software and algorithm analysis. The research model has moved to big data, learning analytics [12], artificial intelligence [13], and information and communication technology [14].

Table 1 main scenarios and technologies of artificial intelligence in education

| Evaluation Aspect | Technological Tools and Methods |
|---|---|
| Student Assessment | Personalized Learning Approaches |
| Examination Scoring & Evaluation | Image Recognition, Computer Vision |
| Personalized Intelligent Tutoring | Data Mining, Intelligent Tutoring Systems, Learning Analytics |
| Smart Classrooms, Smart Schools | Data Recognition, Virtual Labs, Augmented Reality/Virtual Reality |

A classroom is a closed environment where teaching and learning take place. It facilitates the transfer of knowledge from teachers to students and is a crucial component of learning institutions. In a traditional teaching environment, educators can obtain feedback on student learning experiences during face-to-face interactions with students, allowing them to continuously evaluate their instructional plans. Classrooms can be divided into online and offline categories. Online classrooms rely on network technologies, such as signal processing, to facilitate online teaching and school activities. Smart classrooms are a product of online information classrooms and are characterized by the use of information and communication technology. The integration of this technology has had a significant impact on teaching processes and methods, regardless of whether they are traditional offline models or not. Smart classrooms consist of different components based on information technology. Each of these components has its own impact. Collaborate to create a well-designed, interactive, and engaging learning environment. Constructing intelligent classrooms can significantly enhance teachers' instructional abilities, cultivate students' skills, elevate their academic standards, and encourage active participation in the learning process. The use of various technological components, such as interactive boards, cloud management systems, and artificial intelligence computing, can facilitate the delivery of rich content and promote innovative teaching methods that encourage student participation and assessment. This study obtained data on smart classroom research publications from databases obtained from academic search engines such as Google, IEEE, ACM, and CNKI. The presented analysis focuses on a large number of research publications related to intelligence scholarship published between 2000 and 2024. Using a large number of search engines such as Google can provide extensive and detailed search sources. And it can cover computer science, art, humanities, engineering, neuroscience, social science and other fields. Compared with using a single standardized database such as WoS, it can cover a wider range of multidisciplinary topics [15].

This article will review the research directions of smart classrooms from two directions: hardware and software. Hardware includes smart materials and smart environments. Smart materials includes digital materials prepared by teachers for smart classrooms and rich and diverse presentation forms Intelligent physical environment includes intelligent teaching aids and multi-modal display devices. The software mainly includes algorithm analysis and intelligent assessment, mainly including evaluation of students' learning feedback, as well as evaluation of teachers.

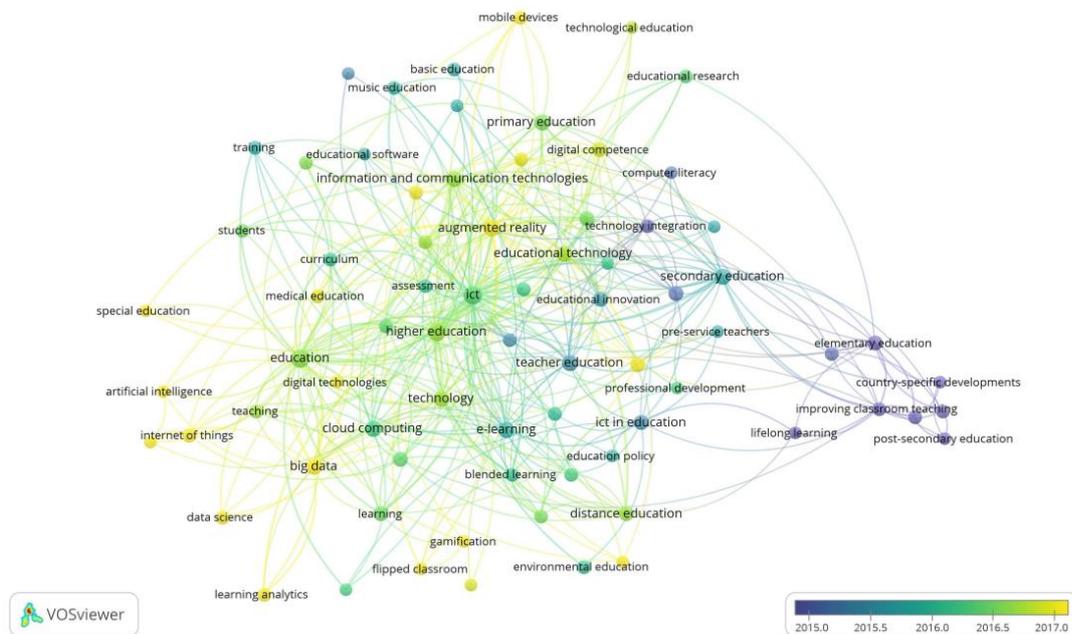

Figure 1 Results of the assessment of technology patterns

## 2.    Hardware

### 2.1.    Traditional offline smart classroom

In traditional offline smart classrooms, hardware is used to support and assist offline teaching. In teaching activities, teachers require two types of materials to support classroom instruction. The first type includes fixed standardized content such as textbooks [16], exam syllabuses [17], and teaching plans, etc. The second type supports the lecture content [18], often more significantly reflecting the teacher's expertise, and is a focal point of smart classroom technology. In traditional classrooms, teachers write or draw on blackboards while students take notes with pen and paper

[19]. In smart classrooms, however, teachers use ICT technologies to display their prepared teaching materials, including presentation documents, films [20], and interactive materials [21], where digital media have replaced notes on the blackboard. These materials allow for the creation of intelligent teaching plans, saving teachers time and making outcomes reusable. Software for creating presentation documents, such as Microsoft PowerPoint, WPS, and Keynote, has become very common. Currently, creative, and interactive options like Prezi and Seewo documents have emerged as new choices for teachers. These web-based platforms also make it easier and more convenient for teacher collaboration and content creation by storing materials in the cloud [22]. Furthermore, sensor devices have been used in preparing teaching materials, such as high-quality cameras [23] and audio recorders [24]. The inclusion of more modal and interactive devices greatly enriches teaching materials and lays the foundation for subsequent modernization of education, information education, and AI-driven education in terms of data volume.

## 2.2. Online smart classroom

The online smart classroom here includes distance education and flipped classroom. These are somewhat different from the traditional offline teaching model. The hardware support of distance education has been completely transformed into a cloud platform, and most of the flipped classroom part relies on online zero-distance inveracities includes the creation of a digital learning environment based on cloud computing [25], a cloud platform tailored for adaptive e-learning [256], and cloud education platforms utilizing VR (Virtual Reality) technology [27]. This kind of research based on platform construction has always been one of the main research directions of educational technology. The development and research of the platform have led to the multi-dimensional growth of teaching plans and materials. At the same time，

The richness of teaching materials and devices leads most directly to the rapid growth of distance learning. Distance learning was once a challenging task in traditional environments, but the digitization of teaching materials and the development of teaching devices, such as interactive whiteboards in virtual classrooms, have significantly improved the efficiency of distance education. Although modern distance education is widely acknowledged as convenient, opinions on its effectiveness vary.

In the context of modern distance education, research on the changes in interaction between teachers and students is framed within three perspectives:

### 2.2.1. Modern distance education weakens the interaction between teachers and students.

According to Li (2020), communication in middle school distance education exhibits three characteristics: (1) the occasional inability for teachers and students to meet, (2) instability in students' emotions and conditions, and (3) difficulty in achieving effective communication between teachers and students within a given time. These aspects contribute to the weakening of effective interaction between teachers and students in distance education [28]. Zhang (2019) also mentions that in an unstandardized learning environment, particularly with the adoption of new educational methodologies, there can be high intensity in learning, low flexibility in timing, and a decrease in motivation and efficiency in learning [29].

### 2.2.2. Modern distance education strengthens the interaction between teachers and students.

Zao (2003) et al. argue that one of the advantages of distance education is the enhancement of interaction. The widespread use of networks strengthens the element of co-learning among students and enables bidirectional interaction online, shifting the educational model from being centered on "receiving education" to focusing on "self-initiated learning". It provides students in different geographical locations with opportunities for learning [30]. Wang (2017) believes that distance education aids in diversifying learning forms, maintaining constant learning motivation, and specifically enables the educational process to be streamlined for efficient learning [31]. Herrington & Kervin (2007) argue that technology in distance education offers opportunities for adopting powerful cognitive tools that students can utilize to solve complex and authentic problems [32].

### 2.2.3. The impact of modern distance education on teacher-student interaction remains unclear.

Li et al. (2018) cautiously concludes in their study on the effect of "Flipped Classroom" teaching on student learning outcomes, based on 37 experiments and quasi-experiments, that flipped classrooms have a moderately positive effect on improving student learning. However, they also note that the flipped classroom approach has a slightly weaker impact on enhancing learning outcomes among younger students. It is challenging to ascertain whether improvements in teacher-student interaction in distance learning, in addition to the interferences of distance lessons, yield a positive influence [33].

## 3. Software

In a broad sense, software refers to the support programs and associated technology that provide in-depth coordination between equipment and materials [34]. It enables an organic combination of these elements, facilitating richer and more expressive teaching methodologies based on planned curriculums. These applications and algorithms enhance content customization and personalization offered by intelligent teaching systems, promoting better absorption and retention of information and improving learners' experience. In the early stages of the development of smart classroom hardware, the development of software and algorithms focused mainly on creating interactive education platforms to improve the quality and efficiency of education. These platforms rely on innovative platforms based on the latest hardware to provide students with a more intuitive and interactive learning experience. For example, the University of Cambridge in the UK has developed GRANTA EduPack, a visual learning software specifically designed for the fields of engineering and design education. GRANTA EduPack enables students to intuitively grasp complex concepts through visual means by providing visually dynamic teaching materials. [35], Meanwhile, in China, researchers and developers have explored an online education architecture based on P2P (peer-to-peer) technology. This architectural design aims to improve the scalability and reliability of the online education platform to cope with the pressure on online servers caused by rapid user growth. This P2P architecture allows users to exchange information directly without relying on a central server, thus achieving greater efficiency and sustainability in terms of network bandwidth and server pressure. [36]. With the perfection of platform construction and the rapid expansion of data, artificial intelligence (AI) assisted education has encountered new opportunities. AI-assisted educational technology can be divided into two main categories: data mining and learning analytics. This section will introduce the software part of smart classrooms based on two major technical categories.

### 3.1. Learning Analytics

3.1 Learning Analytics
Learning analytics focuses on data derived from student characteristics and the establishment of learner models and domain knowledge models. Learning analytics employ technologies related to machine learning, data visualization, learning science, and semantic inference. For instance, AI-based capability learning can generate crucial data from students, offering effective insights into students and predicting key competencies they could pursue, enabling institutions to take proactive measures. In the early stages of AI-assisted educational analysis, multiple-choice questions (MCQs) represented the simplest form of assessment. Moodle [37] and Google Forms [38] were developed to automate the evaluation and statistical analysis of MCQs for students. Subsequently, more modality-based assessment methods emerged, such as an automated system for assessing physics problems developed by Martin, J., which featured a graphical user interface allowing students to present problems and write down answers for automatic grading by the machine [39]. Assessment systems have gradually shifted from visual to textual modalities, with the creation of text-based assessment software, such as the attention-based recurrent convolutional neural network for automatic essay scoring [40], and a multi-scale feature-based essay scoring system [41]. The emergence of these multimodal data-based assessment systems benefits from the vast volume of multimodal information in smart classrooms and, in turn, nourishes the smart classroom systems with AI, better facilitating personalized learning and monitoring student learning situations.

## 3.2. Educational Data Mining

Educational data mining seeks to discover inherent rules and patterns from the vast amount of multimodal data of learners, thereby gaining a better understanding of the educational environment and learners. It mainly includes unsupervised and supervised learning techniques. Unsupervised learning, for example, can analyze demographic data and scoring data from a small number of written assignments [42]. Supervised learning techniques, such as machine learning regression methods, can predict students' future academic performance or dropout risk [43]. Compared to learning analytics, data mining techniques lean more towards knowledge discovery, aiming to uncover hidden data among learners' information, enabling teachers and researchers to make quick adjustments to improve the development of smart classroom systems and further achieve personalized learning. In the early days of smart classrooms, there were many general data mining tools that provided mining algorithms, filtering, and visualization techniques to mine various characteristics of students.

The Early educational data mining work mainly relied on machine learning technologies, which can discover learning patterns and trends by analyzing large amounts of structured data to facilitate the development of smart education. [44]. Over time, rapid advances in the field of Artificial Intelligence have led to more sophisticated algorithms, such as neural networks, which have been successfully integrated into the process of data mining. Neural networks, especially deep learning techniques, have brought unprecedented insights into educational research due to their powerful data processing and non-linear fitting ability capabilities. For example, in the development of English vocabulary learning software researchers from China have expanded from simple vocabulary memorization through deep learning algorithms to how to check and support the state of a student's memory through advanced algorithms. The key advantage of this approach is the ability to customize learning content based on students' individual learning progress and personalized memory profiles, which can greatly improve learning efficiency. By analyzing student interaction data, algorithms based on deep learning can predict which vocabulary is at risk of being forgotten and which words need to be reviewed in a timely manner, thus helping students to consolidate their memories and effectively master the content. [45].

## 4. Conclusion

This paper discusses in depth the development trajectory of the intelligent classroom in the process of education informatization, including the development of hardware facilities from single multimodal to multimodal and the construction of cloud platforms to the current research on software and software algorithms centered on artificial intelligence, covering all aspects of the construction of the intelligent classroom. By combing and analyzing the research literature on smart classroom at home and abroad, it can be seen that with the rapid development of information and communication technology, smart classroom has been gradually transformed from a unimodal, simple information technology application to a complex learning environment that integrates multimodal, advanced artificial intelligence technology. This shift has not only revolutionized teaching methods, but also provided students with a more personalized and efficient learning experience. In future endeavors, the construction and application of smart classrooms will enter a new phase of development as AI and other cutting-edge technologies continue to advance. In particular, the emergence of big models has given the concept of smart teachers a new lease of life. In the future researchers need to work closely with experts in different fields to continuously explore and practice new educational technologies and teaching methods to further explore the reform of education modernization.